\documentclass[sigplan,screen]{acmart}
\usepackage{algorithm,setspace}
\usepackage{adjustbox}
\usepackage[noend]{algpseudocode}
\usepackage[switch]{lineno}
\usepackage{graphicx}
\usepackage{textcomp}
\usepackage{xcolor}
\usepackage{stfloats}
\usepackage{subfig}
\usepackage[T1]{fontenc}
\usepackage[utf8]{inputenc}
\usepackage{lmodern}
\usepackage{booktabs}
\usepackage{listings}
\usepackage[textsize=small,linecolor=white]{todonotes}

\renewcommand\footnotetextcopyrightpermission[1]{} 

\begin{document}
\title{Task-Graph Scheduling Extensions for Efficient Synchronization and Communication
}


 \author{Seonmyeong Bak}
           \affiliation{
           \institution{Georgia Tech}
           \city{Atlanta, GA}
           \country{USA}
           }
           \email{sbak5@gatech.edu}

  \author{Oscar Hernandez}
           \affiliation{
           \institution{Oak Ridge National Laboratory}
           \city{Oak Ridge, TN}
           \country{USA}
           }
           \email{oscar@ornl.gov}

   \author{Mark Gates}
           \affiliation{
           \institution{University of Tennessee, Knoxville}
           \city{Knoxville, TN}
           \country{USA}
           }
           \email{mgates3@icl.utk.edu}

   \author{Piotr Luszczek}
           \affiliation{
           \institution{University of Tennessee, Knoxville}
           \city{Knoxville, TN}
           \country{USA}
           }
           \email{luszczek@icl.utk.edu}

   \author{Vivek Sarkar}
           \affiliation{
           \institution{Georgia Tech}
           \city{Atlanta, GA}
           \country{USA}
           }
           \email{vsarkar@gatech.edu}




\begin{abstract}
Task graphs have been studied for decades as a foundation for scheduling
irregular parallel applications and incorporated in programming models
such as OpenMP. While many high-performance parallel libraries are based
on task graphs, they also have additional scheduling requirements, such
as synchronization from inner levels of data parallelism and internal
blocking communications.

In this paper, we extend task-graph scheduling to support efficient
synchronization and communication within tasks. Our scheduler avoids
deadlock and oversubscription of worker threads, and refines victim
selection to increase the overlap of sibling tasks. To the best of our
knowledge, our approach is the first to combine gang-scheduling and
work-stealing in a single runtime. Our approach has been evaluated on
the SLATE high-performance linear algebra library. Relative to the LLVM
OMP runtime, our runtime demonstrates performance improvements of up to
13.82\%, 15.2\%, and 36.94\% for LU, QR, and Cholesky, respectively,
evaluated across different configurations.



\end{abstract}

\maketitle
\pagestyle{plain}

\section{Introduction}
\label{sec:intro}
On-node parallelism in high-performance computing systems has increased
significantly over the past years.
This massive amount of parallelism has the potential to
deliver significant speedups, but there is a  concomitant burden on
application developers to exploit this parallelism in the presence of
inherent load imbalances and communication/synchronization requirements.
One popular approach to reduce the complexity of application
development for modern processors is to introduce high-performance
libraries.  High-performance linear algebra libraries have pioneered the
use of task graphs to deal with load imbalances in
parallel kernels such as LU, QR, and Cholesky factorizations while also
exploiting data locality across dependent blocks.

At the same time, there is now increased support for task-parallel
execution models with task dependencies in modern parallel programming
models, such as OpenMP.
Many task graphs in real-world applications include library
calls or nested  parallel regions that involve blocking operations such
as barriers. They often include mixed sequences of communication and
computation operations for latency hiding. The tasks also often create
groups of child tasks to exploit potential available resources. However,
 current task-based programming models are unable to support  these real-world application requirements, which motivates the work
presented in this paper.

\begin{figure}[b]
\captionsetup[subfloat]{farskip=2pt,captionskip=1pt}
  \subfloat[Deadlock in nested parallel regions of tasks or ULTs] {
  \includegraphics[width=0.47\textwidth, trim=0 0 0 5]{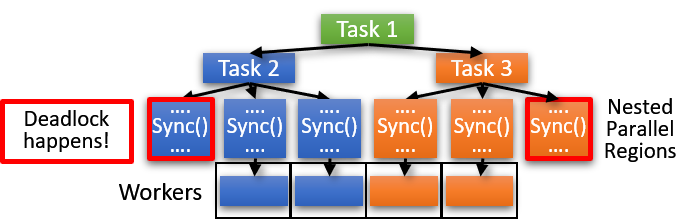}
  \label{fig:nested_par_intro}
  }\\
  \subfloat[Deadlock avoidance with gang-scheduling of nested parallel regions]{
     \includegraphics[width=0.42\textwidth, trim=0 5 0 5]{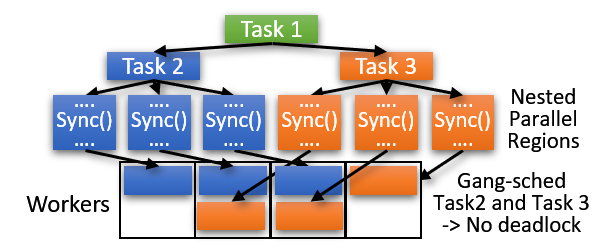}
     \label{fig:gang_sched_intro}
  }
  \vspace{-2mm}
  \caption{Deadlock issues in nested parallel regions from a group of tasks or User-level Threads(ULT)}
\end{figure}

Further, tasks often spawn nested parallel regions through calls to
library functions or  user code with internal parallelism. These nested
parallel regions lead to execution by additional pools of threads,
which in turn causes oversubscription of cores. This oversubscription
can delay  intra/inter-node communication or synchronization operations,
which often occur in periodic time steps. Scheduling these operations
without interference from other parallel regions helps reduce the
overall critical path of the application. On the other hand, delaying
the execution of communication operations can lead to overall degraded
performance.
One approach to addressing the challenge of oversubscription in nested
parallel regions is to
adopt the use of user-level threads
(ULTs)\cite{BOLT_PACT19,ccgrid2018:seon}. However, ULTs cannot support
general nested parallel regions involving blocking synchronization and
communication operations.  In general, adopting ULTs can lead to
deadlock because all of the ULTs are not guaranteed to be scheduled onto
worker threads when a blocking operation occurs.
Figure~\ref{fig:nested_par_intro}(a) shows how adopting ULTs can lead to
deadlock when a nested parallel region contains blocking synchronization
operations.

In this work, we show how a standard task scheduling runtime system can
be extended to support the real-world constraints discussed above by (1)
combining gang-scheduling and work-stealing and (2) supporting hybrid
victim selection.  Our approach provides deadlock-avoidance in the
scenario where multiple user-level contexts are synchronized with
blocking operations.
The integration of gang-scheduling with work-stealing helps nested
parallel regions run efficiently without oversubscription and deadlock.

The parallel regions to be gang-scheduled are created as ULTs and
scheduled onto a consecutive set of cores that are close to the worker
that executed the task that initiated the parallel region, as shown in Figure~\ref{fig:nested_par_intro}(b). Workers can
schedule other tasks in work-stealing mode while they are
gang-scheduling ULTs from specified parallel regions. ULTs that are
gang-scheduled on reserved workers can steal tasks from their parallel
region when they reach a join barrier. When multiple gangs are created
within the same node, they're ordered globally to prevent a deadlock
across ULTs that are scheduled on workers. This hybrid scheduling of
gang-scheduling and work-stealing reduces interference and increases
data locality for data parallel tasks that involve synchronization and
communication in each time step.

\begin{figure}[h]
\vspace{-1mm}

  \includegraphics[width=0.46\textwidth, trim=0 15 0 15]{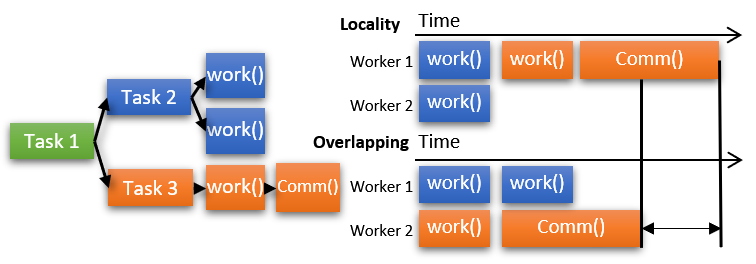}
 \vspace{-2mm}
  \caption{Difference in critical path of mixed sequences of communication and computations}
  \label{fig:victim_select}
  \vspace{-4mm}
\end{figure}

In addition to  gang-scheduling,
our runtime system adopts a hybrid victim selection policy in
work-stealing to increase communication-computation overlap
as well as data locality. Figure~\ref{fig:victim_select} shows the
performance difference from different victim selection policies. The
existing OpenMP runtime systems schedule tasks as in the \emph{Locality} case,
while our approach pursues both the \emph{Locality} and the \emph{Overlapping} cases. To the best of our knowledge,
ours is the first work to propose and implement a hybrid scheduling of
gang-scheduling and work-stealing as well as hybrid victim selection in
a production-level runtime system and to demonstrate the implementation
on real-world examples.

The contributions of this paper are as follows:
\begin{itemize}
    \item Extension of task-based runtime systems to integrate
          gang-scheduling with work-stealing in an efficient manner.
    \item Introduction of hybrid victim selection to increase the
          overlap of tasks in task graphs while still preserving data
          locality.
    \item Evaluation of our approach on real-world linear algebra
          kernels in the SLATE library: LU, QR, and Cholesky
          factorizations.  Relative to the LLVM OMP runtime, our runtime
          demonstrates performance improvements of up to 13.82\%, 15.2\%, and
          36.94\% for LU, QR, and Cholesky, respectively, evaluated across
          different configurations.
\end{itemize}

\section{Background}
\subsection{Task graphs in Task-Level Programming Models}

Many task-level parallel programming models have introduced task graphs
in different ways to extract parallelism from irregular parallel
applications. The first type of interface for task graphs is
\emph{explicit task dependency} through objects such as promises and
futures in C++ 11~\cite{C++11} and Go~\cite{go}. Tasks wait on
objects until the predecessors of the objects put data on the
objects, which resolve the dependencies of the successors. The other
type is \emph{implicit task dependency}, which automates the management of
objects to improve programmability with the help of compiler and runtime
systems that form dependencies through directives as \emph{depend} in OpenMP
4.0~\cite{openmp4.0} or data flow of variables. After dependencies of
tasks are resolved, they become \emph{ready tasks} and are treated as
normal tasks. Most task-based runtime systems including OpenMP use
per-thread stealing queues so threads where the tasks become ready push
tasks to their local work-stealing queue.



\subsection{User-level threads for Task-Level Programming Models}


In parallel programming models, user-level threads (ULTs) have been used
to resolve oversubscription issues by scheduling user-level threads onto
kernel-level threads (KLTs) when multiple parallel regions are running on
the same cores. The mapping of ULT to KLT enables lightweight context
switching through storing necessary data for context switching in a user space rather than in a kernel space.
There have been
several implementation of user-level threads to benefit from its
lightweight context switching in different contexts~\cite{converse,
qthread, argobot}. In spite of the benefits of ULT, they have deadlock
issues because of a lack of coordination with kernels as described in
Figure~\ref{fig:nested_par_intro}(a). The OS kernel cannot identify the status
of each ULT, which can lead to deadlock if user-level threads encounter
blocking operations such as barriers and locks. There have been several
efforts where runtime systems share ULT information with the OS kernel, such as scheduler activations~\cite{sched_act}.
However, the previous works require significant changes in both the ULT runtime and
OS kernel, which has inhibited  the adoption of their APIs in
operating systems.



\subsection{Gang-scheduling and Work-stealing}


Gang-scheduling~\cite{gang_sched_JPDC, gang_sched_ICDCS} was initially
proposed to reduce the interference of a group of threads by other threads or processes. Gang-scheduling, as first
introduced, uses a matrix to pack thread requests from processes in
which each row is scheduled one at a time. Thus, context switching
occurs when it moves from one row to the next row, which reduces the
delay in communication across threads incurred by unnecessary context
switching. However,  a waste of resources results when the threads in
each gang have a load imbalance or insufficient cores are available to
meet their requests. Different packing policies have been proposed to
address these inefficiencies~\cite{gang_sched_JPDC,gang_sched_pack,paired_gang_sched}, but
they did not solve the issue completely. Also, gang-scheduling introduces
significant overhead through its use of global data structures.
In contrast, work-stealing is a distributed scheduling policy in which
each worker schedules tasks independently. Each worker creates tasks and
pushes them into their work-stealing queues. Then, other workers steal
tasks from the worker by running a work-stealing algorithm
independently. Work-stealing maximizes load balancing but incurs overheads due to reduced locality through context switching as well as communication delays, since locality and communication are not part of the task-parallel models that work-stealing schedulers were designed to support.  Extended work-stealing algorithms have
been introduced to alleviate the cost of work-stealing by considering
the locality of participating processing elements~\cite{SPAA00_Umut,
guo2010slaw, SC14_lifflander}. Some of the previous works also extended
work-stealing to distributed
systems~\cite{SC09_ScalaWS,HPDC12_lifflander, Euro-Par10_HWS}.

\section{Design}
This section describes the algorithm and interface we designed to
address the limitations of current task-parallel runtimes mentioned in
Section~\ref{sec:intro}. We propose the use of {\em gang-scheduling} to
schedule ULTs of a parallel region without oversubscription and
deadlock.  Our design supports the
use of gang-scheduling for specific parallel regions or globally,  while
other parallel regions and tasks are scheduled with work-stealing. In
addition to gang-scheduling, we also discuss how the victim selection
policy, which impacts how a task graph is traversed, affects the
overlapping of communication and computation tasks, and we propose a
{\em hybrid victim selection policy} to improve the overlapping
supported by the task scheduler.


\subsection{Gang-Scheduling of Data-Parallel Tasks}

\subsubsection{Integrating gang-scheduling with work-stealing}

Gang-scheduling and work-stealing have been thought of as oil and water
in task scheduling. Each has its advantages and disadvantages as
compared to the other. Integrating them so that each can be used in
cases when it is beneficial can help improve the overall performance of
task-parallel applications. We propose extending the \emph{omp parallel}
construct to schedule threads of selected parallel regions in
gang-scheduling mode. Users can apply gang-scheduling to upcoming or all
parallel regions through our proposed API in
Listing~\ref{code:gang_api}. By default, all top-level parallel regions
are scheduled in gang-scheduling mode. Other parallel regions that are
not set by the proposed API are scheduled in work-stealing mode by
putting all their ULTs into the calling worker's local work-stealing
queue. For the rest of this paper, we refer to ULTs to be scheduled in
gang-scheduling mode as \emph{gang} ULTs, while other ULTs and tasks are
referred to as \emph{normal} ULTs and tasks.
\begin{lstlisting}[captionpos=b, caption={API to apply gang-scheduling to parallel regions}, label=code:gang_api]
export OMP_GANG_SCHED=1; //Apply gang-scheduling to all parallel regions
void ompx_set_gang_sched(); // All following parallel regions are gang-scheduled after this call
void ompx_reset_gang_sched(); // Parallel regions after this call are scheduled in default scheduling policy
\end{lstlisting}



\begin{algorithm}[b]
\footnotesize
\setstretch{0.9}
\begin{algorithmic}[1]
\Function {gang\_sched}{$n\_request, threads$}
\State \Comment{Gang-schedule \emph{threads} to \emph{n\_request} workers}
\State $gang\_id \gets monotonic\_gang\_id()$
\State $workers \gets get\_workers(n\_request)$
\State $n\_gang\_threads \gets n\_gang\_threads + n\_request$
\For{$i = 0\ to\ n\_request-1$}
\State $thread[i]_\mathrm{gang\_id} \gets gang\_id$
\State $thread[i]_\mathrm{nest\_level} \gets cur\_worker_\mathrm{nest\_level}$
\State $push\ thread[i]\ to\ worker[i]_\mathrm{gang\_deq}$
\EndFor
\EndFunction

\Function {get\_workers}{$n\_request$}
\State \Comment{Retrieve a list of least loaded \emph{n\_request} workers}
\State $avg\_load \gets n\_gang\_threads / n\_workers$
\If {$cur\_worker\_id + n\_request >= n\_workers$}
\State $start\_worker \gets cur\_worker\_id - n\_request/2$
\Else
\State $start\_worker \gets cur\_worker\_id+1$
\EndIf
\State $idx \gets start\_worker, i \gets 0$
\While {$i < n\_request$}
\If {$worker[idx]_{n\_gang\_threads} <= avg\_load$}
\State $reserved\_workers[i++] \gets worker[idx]$
\EndIf
\State $idx \gets (idx +1) \% n\_workers$
\EndWhile
\Return $reserved\_workers$
\EndFunction

\Function {is\_eligible\_to\_sched}{$thread$}
\State \Comment{Check if \emph{worker} can steal \emph{thread}}
\If{$worker_{cur\_gang\_id} < 0$}
    \Return true
\EndIf
\If{$thread_{nest\_level} > worker_{nest\_level}$} 
    \Return true
\ElsIf{$thread_{nest\_level} = worker_{nest\_level}$ \par
    \hspace*{1em}$\land \ thread_{gang\_id} < worker_{gang\_id}$}
\Return true
\EndIf
\State
\Return false
\EndFunction
\caption{Gang-scheduling with Load Balancing and Deadlock Avoidance}
\label{alg:gang}
\end{algorithmic}
\end{algorithm}

\subsubsection{Gang-scheduling of user-level threads without deadlock}

When multiple gang-scheduled parallel regions are running
simultaneously, it is important they be scheduled without the
possibility of deadlock. To prevent deadlock as described in
Figure~\ref{fig:nested_par_intro}, we assign a monotonically increasing
\textbf{\emph{gang id}} to each parallel region, which is incremented
atomically across all workers. We use this \emph{gang id} to restrict
the scheduling order of gangs so as to guarantee that deadlock does not
occur. Algorithm \ref{alg:gang} describes how the \emph{gang} ULTs from
a parallel region are assigned the \emph{gang\_id} and
\emph{nest\_level} of the current worker; the runtime system then
gang-schedules \emph{gang} ULTs of each parallel region.
\textbf{\emph{gang\_sched()}} is synchronized by a shared lock in the
\emph{fork} stage of a region in the OpenMP runtime. The \emph{fork}
phase involves access to global data structures which are synchronized
by a global lock for the \emph{fork} and \emph{join} phases in the
runtime system. Thus, parallel regions have an inevitable serialization
in the \emph{fork} phase, and \emph{gang\_sched} contributes a marginal
additional waiting time to the \emph{fork} phase of each region.

When each gang is assigned a set of workers (``reserved'' workers), the
number of gang ULTs and the distance of each worker from the master
thread are considered. We assume that all the worker threads are pinned
to avoid any migration cost and uncertainty that may be caused by the OS
thread scheduler.
The workers that are closer to the current worker and less loaded with
gang-scheduled ULTs have higher priority in
\textbf{\emph{get\_workers()}}. Workers are selected to be as close to
each other (preferably, consecutive) as possible.

\emph{Gang} ULTs become stealable after they are scheduled onto the
reserved workers. Other workers can steal the \emph{gang} ULTs from the
reserved workers, which enables an earlier start of \emph{gang} ULTs if
the reserved workers for the gang are busy executing other \emph{normal}
ULTs and tasks. This is because we only consider the number of
\emph{gang} ULTs on each worker. This additional work-stealing resolves
unidentified load imbalance without tracking all \emph{normal} ULTs and
tasks.
The work-stealing of \emph{gang} ULTs happens at every scheduling point,
such as barriers, along with \emph{normal} tasks and ULTs. \emph{Gang}
ULTs have the highest priority in work-stealing and go through an
additional function to check if each \emph{gang} ULT from a victim
worker can be scheduled on the caller through
\textbf{\emph{is\_eligible\_to\_sched}}. This function compares the
\emph{nest-level} and \emph{gang\_id} of the current worker with the
corresponding variables in the victim \emph{gang} ULT which are assigned
in \textbf{\emph{gang\_sched}}.
This function guarantees parallel regions are scheduled in a certain
partial order where gangs, which are started earlier or in lower nested
levels, have precedence over those that started later or are in upper
levels.
In this way, our gang-scheduling approach prevents deadlock of multiple
parallel regions contending on the same pool of workers as described in
Figure~\ref{fig:nested_par_intro}, allowing us to benefit from
work-stealing for load balancing.

When \emph{gang} ULTs reach a join-barrier at the end of a parallel
region, they can steal \emph{normal} ULTs and tasks from workers in
parallel regions of the upper nest level even when they're not reserved
for the gang. When any stolen task spawns a parallel region, the task is
suspended to prevent a waiting time incurred by the new nested parallel
region. Each suspended task is pushed back to a separate work-stealing
queue for suspended tasks. These tasks have higher priority than other
tasks.

\subsubsection{Comparison with previous work}

With the algorithms and heuristics described in this section, only
selected parallel regions are guaranteed to be scheduled in
gang-scheduling mode. The gang-scheduling we proposed is relatively
relaxed compared with previous work because our algorithm guarantees a
parallel region to run simultaneously at some point in runtime. Some of
the threads in the region can run earlier than others, which may lose
the locality of  stronger approaches to gang-scheduling. However, this
relaxed gang-scheduling algorithm can also result in less waiting time
and more efficient use of workers. Our scheme doesn't require a global
table to keep track of threads and reduces waiting time by allowing each
region to start immediately and to make ULTs stealable after being
gang-scheduled.

\subsection{Hybrid Victim Selection for Overlapping and Data Locality}

Task graphs involving communication and computation tasks are commonly
used to exploit parallelism by overlapping tasks in different iterations
of iterative applications. In linear algebra kernels, block-based
algorithms have similar task graphs to overlap the waiting time of
current tasks by doing some computation for the next tasks.
As mentioned in Section~\ref{sec:intro}, many task-level runtime systems
use heuristics to schedule tasks in task graphs to maximize data
locality. One of the common heuristics is to use a history of previous
successful steals. This heuristic is intuitively helpful for data
locality by making workers steal the same loaded victim threads until
their task queue becomes empty. However, this heuristic may prevent the
overlapping of communication and computation across sibling tasks.
When one task becomes ready earlier than another and both of them have
nested child tasks to exploit potential available parallelism, the
history-based heuristic makes all workers first steal the child tasks
from the first task, before moving on to the next task---even though the next
task becomes ready while the first and its child tasks are being
executed. This prevents overlapping of communication on the next task
with computation on the first task.
\begin{algorithm}[b]
\footnotesize
\setstretch{0.9}
\begin{algorithmic}[1]
\Function {Do\_WorkStealing}{}
\State \Comment{Steal a task from other workers, record the steal for the next steal}
\State $cur\_idx \gets cur\_worker_\mathrm{history\_idx}$
\State $victim \gets select\_victim(),\  task \gets steal\_task(victim)$
\If {$task\ is \ valid$}
\State $cur\_worker_\mathrm{prev\_victim\_id[cur\_idx]} \gets victim$
\State $increment\ cur\_worker_\mathrm{history\_idx}$
\Else
\State $cur\_worker_\mathrm{prev\_victim\_id[cur\_idx]} \gets -1$
\State $decrement\ cur\_worker_\mathrm{history\_idx}$
\EndIf
\State \Return $task$
\EndFunction
\Function{Select\_Victim}{}
\State \Comment{Retrieve a worker id from previous steals or rand()}
\State $cur\_idx \gets cur\_worker_\mathrm{history\_idx}$
\If {$cur\_worker _\mathrm{prev\_victim\_id[cur\_idx]} >= 0$}
\State $victim\_id \gets cur\_worker _\mathrm{prev\_victim\_id[cur\_idx]}$
\Else
\State $victim\_id \gets rand()\ \%\ n\_workers$
\EndIf
\Return $victim\_id$
\EndFunction
\caption{Work-Stealing with hybrid of history and random victim selection}
\label{alg:victim_choose}
\end{algorithmic}
\end{algorithm}
To resolve these unintended anomalies, we tested random work-stealing,
which just chooses a victim thread randomly without the use of history.
This random stealing, however, suffers from a loss of data locality.
Thus, we combined  history-based and random work-stealing so that each
worker alternatively steals tasks from its history of successful steals
and from random victims. This simple heuristic can make use of data
locality and overlapping of communication and computation tasks.

Algorithm \ref{alg:victim_choose} is a combined algorithm that chooses
victim workers for stealing. Each worker calls \emph{do\_workstealing}
when their local-task queue is empty and waiting for other threads on
any synchronization point. First, each worker tries to retrieve the
victim thread from its local history of steals. If this steal turns out
to be successful, then it moves to the next slot in the local steal
history array. This makes the worker try random-stealing after a
successful steal. If the current steal fails, regardless of whether it
uses history or a randomly chosen victim, it moves back to its previous
slot in the history array. If the entry has a valid victim thread id,
this worker will try to steal from the victim where the latest
successful steal occurred. If not, it keeps stealing from randomly
chosen victims. This combined selection of victim from history and
random method prevents workers from repeatedly stealing from the same
victim, which would result in a serialized sequence of communication and
computation without overlapping.

\section{Implementation}
\label{sec:impl}
In this section, we introduce our integrated runtime system of
Habanero-C library and LLVM OpenMP runtime to implement the proposed
gang-scheduling algorithm and victim selection policy.

\subsection{Overview of Our Implementation}
\begin{figure}[b]
    \includegraphics[width=.47\textwidth, trim=0 10 0 20]{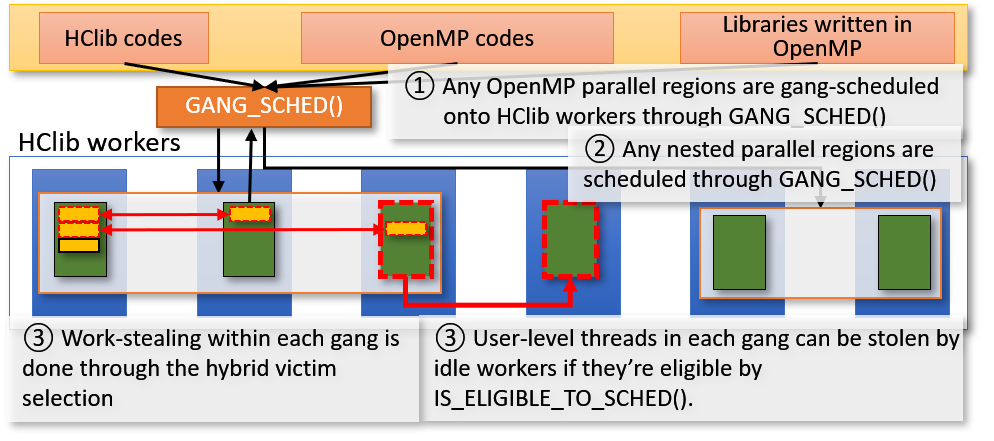}
    \vspace{-2mm}
    \caption{Implementation of Integrated HClib and OpenMP runtime}
    \label{fig:overview}
\end{figure}
We integrated LLVM OpenMP runtime and Habanero-C library (HClib) to use
HClib's user-level threading routines. This integrated runtime creates
OpenMP threads as user-level threads that run on HClib workers. This
runtime can run pure C++ codes using HClib APIs, OpenMP codes, and HClib
with OpenMP codes. In this work, we use pure OpenMP codes to focus on
the task dependency graph issues in production-level applications. The user
needs to load this library to their application binary using OpenMP
through \emph{LD\_PRELOAD}. The LLVM OpenMP runtime supports gcc, icc, and
clang, so any OpenMP binary built with the compilers can run on our
integrated runtime without any change to their codes.

Figure~\ref{fig:overview} shows how OpenMP instances are scheduled onto
HClib workers when gang-scheduling is enabled through the interface in
Algorithm~\ref{code:gang_api}. User-level threads in each gang can be
stolen by idle workers. When idle workers try to steal a ULT from any
gang, they check with \emph{IS\_ELIGIBLE\_SCHED} function if it is fine
to schedule the ULT by comparing their active \emph{gang\_id} and
\emph{nest\_level} with the ULT. Within each gang, OpenMP threads steal
tasks through the hybrid victim selection. In the following sections,
we will describe how we implement gang-scheduling and work-stealing for
nested-parallel regions in this integrated runtime system.

\subsection{Scheduling of Parallel Regions on the shared pool of workers}

Multiple OpenMP instances can run on this integrated runtime system by
gang-scheduling and work-stealing, so workers may have different
nest-levels. User-level threads from each OpenMP instance running on the
workers should be able to get access to each other. So, we implemented
that each worker has arrays for its active \emph{gang\_id},
\emph{nest\_level} and \emph{thread\_array}. These arrays are indexed by
\emph{internal\_nest\_level} of each worker to point to an active entry
for the current running parallel region.

\begin{figure}[b]
\vspace{-3mm}
    \centering
    \includegraphics[width=.47\textwidth,trim=0 0 0 10]{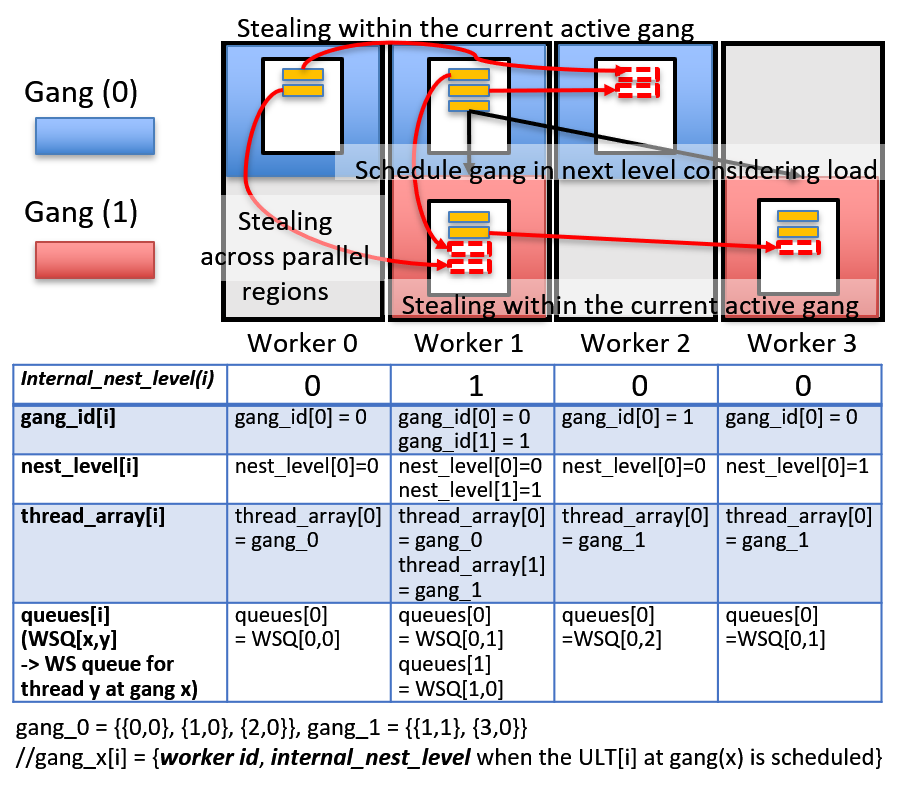}
    \vspace{-2mm}
    \caption{Gang-scheduling for nest-parallel regions and Work-stealing within and across gangs}
    \label{fig:nest_par_impl}
\end{figure}

Figure~\ref{fig:nest_par_impl} shows how our implementation schedules
multiple parallel regions onto the shared workers. When any ULT on each
worker tries to schedule a new OpenMP instance onto workers, it creates a
new \emph{thread\_array} which is assigned an atomically incremented
\emph{gang\_id}. Each ULT also contains a copy of \emph{gang\_id},
\emph{nest\_level} and pointer to \emph{thread\_array}.
When each ULT is eligible on a worker by \emph{IS\_ELIGIBLE\_SCHED},
it is stolen by the worker, which copies the information of the ULT to
its local entries indexed by \emph{internal\_nest\_level}
for \emph{gang\_id}, \emph{nest\_level} and \emph{thread\_array}. The
worker store its \emph{worker id} and \emph{internal\_nest}
\emph{\_level} in the \emph{thread\_array[internal\_nest\_level][the
ULT's thread id]} where other ULTs can find the ULT and its
work-stealing queue on this worker. So, other workers scheduling ULTs in
the same OpenMP instance steal a task through this shared
\emph{thread\_array}. Each worker keeps a separate array of queues for
\emph{normal} ULTs and tasks indexed by \emph{internal\_nest\_level},
which are reused without being reallocated for each new instance. For
\emph{gang} ULTs, each worker has a local \emph{gang\_deq} where a
master thread initiating a parallel region pushes a \emph{gang} ULT
through \emph{gang\_sched} function in Algorithm~\ref{alg:gang}, which
has highest priority over other queues. Each worker gets a ULT by
atomically popping from this \emph{gang\_deq}. On any scheduling point,
each worker checks this queue first before they schedule tasks in
\emph{queues[internal\_nest\_level]}.

\subsection{Work-stealing across different Parallel Regions}

Each gang has reserved workers. Any synchronizations, such as barriers and
locks, are handled without deadlock within each gang. Each worker does
work-stealing among workers where ULTs in the same gang are running. As
mentioned above, each worker finds a work-stealing queue of a victim ULT
through recorded info in the shared \emph{thread\_array}. Work-stealing
across different parallel regions is not allowed in the middle of each
parallel region. When each ULT reaches its join barrier at the end of
its parallel region, it can steal tasks from other parallel regions.
This work-stealing out of parallel regions is allowed because we assume
there is no work left until the end of this parallel regions, and this
cross-region work-stealing has been proven to help reduce the idle time
of unbalanced parallel regions in previous works~\cite{ccgrid2018:seon}.
If any stolen task leads to a nested parallel region, the task is
suspended and pushed to the worker's local work-stealing queue for
suspended tasks, which has the highest priority over other queues. To
prevent any possibility that the work-stealing can lead to a deadlock by
creating a cycle, we restrict this out-of-region work-stealing to happen
from lower nested parallel regions to upper parallel regions on each
worker. In other words, each worker can do this out-of-region stealing
at \emph{thread\_array[internal\_nest\_level:0]}.

\section{Application Study}
We use three
linear algebra kernels from the SLATE library~\cite{SLATE_SC19} to
showcase the benefits of our work: LU, QR, and Cholesky.
 SLATE is a state-of-the-art library developed by the University of
Tennessee that is designed to make efficient use of the latest multicore CPUs and
GPUs in large-scale computing with common parallel computing techniques
such as wavefront parallelism for latency hiding and heterogeneous use
of CPU and GPU in distributed environments. 
SLATE outperforms existing vendor-provided libraries and
its predecessor, ScaLAPACK~\cite{ScaLAPACK}.
For our evaluation, we used the NERSC Cori GPU cluster and built SLATE from its main
repository\footnote{https://bitbucket.org/icl/slate} with the configuration in
Table~\ref{tab:config}.
For the baseline OpenMP runtime system, we used the LLVM OpenMP runtime,
which was forked from the LLVM github repository on 06/29/2020.

\begin{table}[h]
\vspace{-2mm}
\renewcommand{\arraystretch}{1.0}
\normalsize
\begin{adjustbox}{width=0.475\textwidth, center}
\begin{tabular}{@{}l|l|ll@{}}
\toprule
\multicolumn{2}{c|}{\textbf{Hardware Configuration (per node)}} & \multicolumn{2}{c}{\textbf{Software Configuration}}        \\ \midrule
\textbf{Cluster}      & NERSC Cori GPU                          & \multicolumn{1}{l|}{\textbf{SLATE}}    & 06/22/2020 Commit \\
\textbf{CPU}          & 2 x Intel Skylake 6148 (20C, 40SMT)     & \multicolumn{1}{l|}{\textbf{Compiler}} & GCC 8.3           \\
\textbf{GPU}          & 8 x Nvidia V100                         & \multicolumn{1}{l|}{\textbf{MKL}}      & 2020.0.166        \\
\textbf{NIC}          & 4 x dual-port Mellanox EDR              & \multicolumn{1}{l|}{\textbf{CUDA/MPI}}      & 10.2.89, OpenMPI 4.0.3     \\ \bottomrule
\end{tabular}
\end{adjustbox}
\caption{Hardware/Software Configuration for Experiments}
\label{tab:config}
\vspace{-5mm}
\end{table}

We tested different configurations of ranks-per-node and cores-per-rank using the LLVM OMP baseline, and selected the best configurations for all our experiments as follows.
For LU and QR,
we ran each kernel with 4 MPI ranks on each node with 10 OpenMP threads per rank
, while for Cholesky, we used 2 MPI ranks per node with 20 OpenMP threads per
rank. For GPU runs, we used 4 GPUs per node which showed the
best baseline performance.
The OpenMP threads and HClib workers are pinned in the same fashion, using the best affinity setting among those tested.

We ran SLATE's performance test suite to measure the performance of
each kernel in GFlops with different configurations. Each performance
measure is a mean of 6 runs after the first run as warm-up.
We ran the kernels with small and large matrices to cover common sizes
of input matrices on single and multi-node runs. For GPU runs, we used
only large matrices where the GPU version starts to outperform the
CPU-only runs.
For Cholesky, we ran the CPU-only version because the GPU
version of Cholesky offloads the trailing matrix update to the GPU, without offering an opportunity to overlap the trailing task and panel task (since no prior runtime was able to exploit this overlap using the victim selection approach in our runtime).

For comparison, we ran the test suite with the ScaLAPACK reference implementation
using sequential MKL (denoted by
\emph{ScaLAPACK (MKL)}), the SLATE default implementation using \emph{omp
task depend} on LLVM OpenMP runtime (denoted by \emph{LLVM OMP}), and
the same SLATE implementation on our integrated runtime (denoted by
\emph{HClib OMP}).


\subsection{Overview of Task Graphs for LU, QR, and Cholesky in SLATE}

\begin{figure}[t]
    \centering
     \includegraphics[width=0.35\textwidth, trim=0 10 0 10]{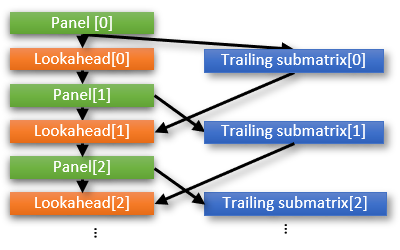}
    \caption{Simplified task graph of factorization kernels in SLATE}
    \label{fig:task_graphs}
    \vspace{-4mm}
\end{figure}

Figure \ref{fig:task_graphs} shows the general form of task graphs for
factorization kernels in SLATE. SLATE uses lookahead tasks and panel
factorization for overlapping of computation and communication as well
as data locality. Factorization kernels factor panels (each
panel is a block column) and then send tiles in the factored
panel to other ranks so that they can update their next block column and
trailing submatrix. Lookahead tasks update the next block column for
the next panel factorization, and the trailing submatrix task updates
the rest of the trailing submatrix. Panel and lookahead tasks are assigned a higher
priority than trailing submatrix computation with a \emph{priority}
clause to accelerate the critical path of the task graph, which is
supported by only a few OpenMP runtime systems such as GNU OpenMP.
Regardless of the support of \emph{priority}, it doesn't guarantee that
the scheduling of higher priority tasks will precede lower priority
tasks even when it is supported because a \emph{priority} clause simply
gives precedence to only \emph{ready tasks} specified with higher
priority. The \emph{trailing submatrix[i-1]} task and its child tasks
become ready earlier than the \emph{panel task[i]} and its child tasks.
For this sequence of tasks, the common history-based work-stealing can
prevent the expected overlapping of computation in \emph{trailing
submatrix} and communication in \emph{panel task}.
Cholesky factorization has significant degradation from this anomaly.

Each factorization kernel has a different series of computations and
communication routines in the panel, lookahead, and trailing submatrix
tasks depending on its algorithm. Each of the tasks consists of a block of
columns. In the following sections, we'll discuss in detail how our
suggested approaches improve the performance of these kernels.

\subsection{LU, QR Factorization: Gang-Scheduling of Parallel Panel Factorization}

\begin{figure}[b]
\vspace{-2mm}
    \centering
     \includegraphics[width=0.4\textwidth, trim=0 20 0 10]{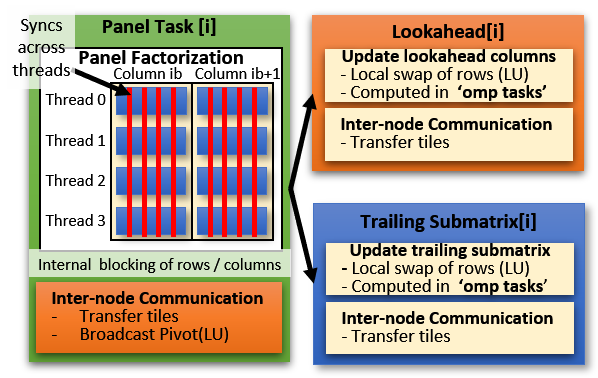}
    \caption{Panel, lookahead, and submatrix tasks of LU and QR in SLATE}
    \label{fig:lu_qr_detail}
\end{figure}
LU factorization is a basic factorization kernel for solving linear
systems of equations in which the coefficient matrices are
non-symmetric. Several optimizations for LU factorization have
been suggested. SLATE adopts a multi-threaded panel algorithm to achieve
a best-performing LU implementation~\cite{LU_Chol_SLATE}.
Figure~\ref{fig:lu_qr_detail} shows what each task in the task graph in
Figure~\ref{fig:task_graphs} does in the LU and QR factorization of SLATE.
First, the LU factorization in SLATE does a panel factorization on a block
of columns in panel tasks. The panel factorization is parallelized in a
nested-parallel region.

\begin{figure}[b]
\vspace{-3mm}
\footnotesize
\captionsetup[subfloat]{farskip=2pt,captionskip=1pt}
  \subfloat[LU, 1-node, small size] {
    \includegraphics[width=.22\textwidth, trim=10 5 15 5]{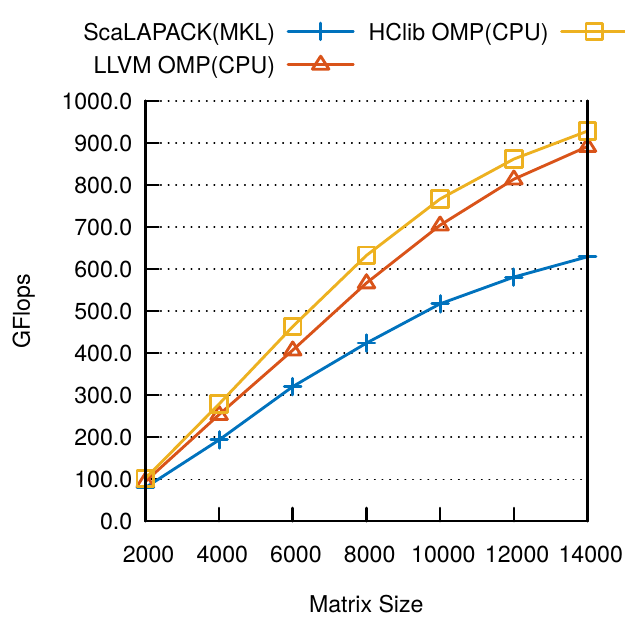}
    \label{fig:perf_lu_small}
  }\hfill
  \subfloat[LU, 1-node, large size] {
  \includegraphics[width=.22\textwidth, trim=10 5 15 5]{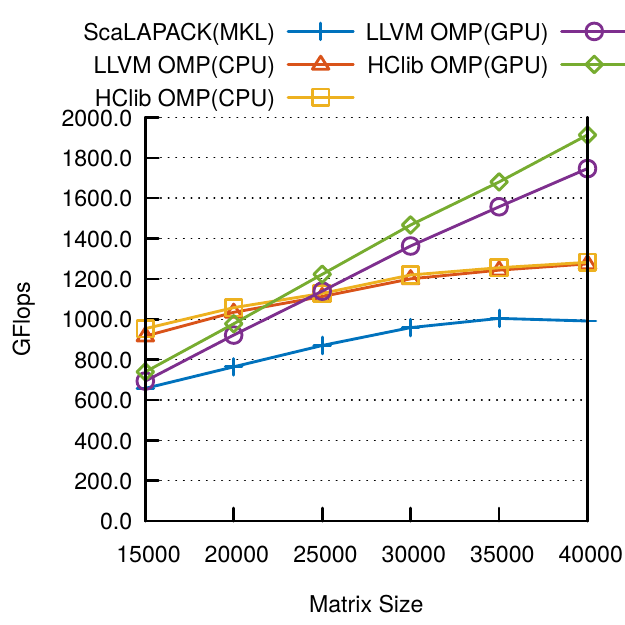}
  \label{fig:perf_lu_large}
  }\vfill
  \subfloat[QR, 1-node, small size] {
    \includegraphics[width=.22\textwidth, trim=10 5 15 5]{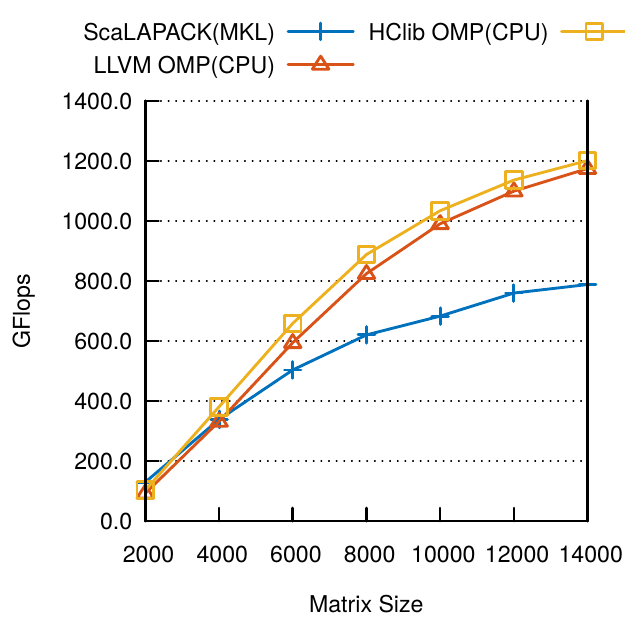}
    \label{fig:perf_qr_small}
  }\hfill
  \subfloat[QR, 1-node, large size] {
  \includegraphics[width=.22\textwidth, trim=10 5 15 5]{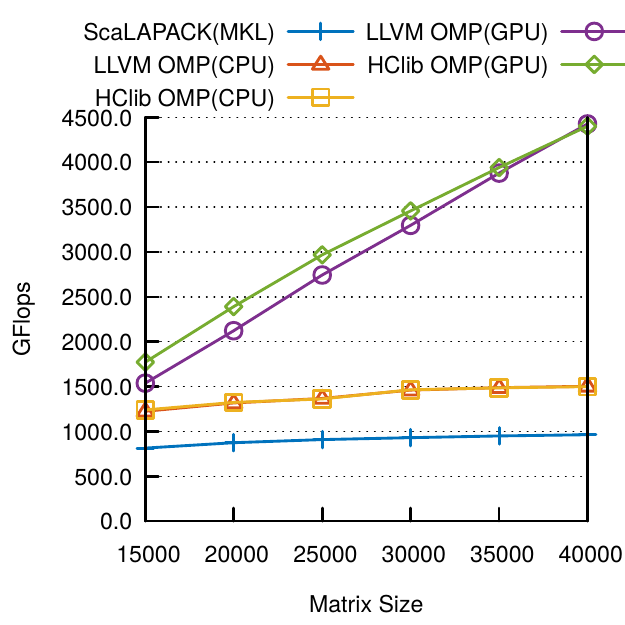}
  \label{fig:perf_qr_large}
  }\vfill
  \subfloat[LU, 4-node, large size]{
    \includegraphics[width=.22\textwidth, trim=10 5 15 5]{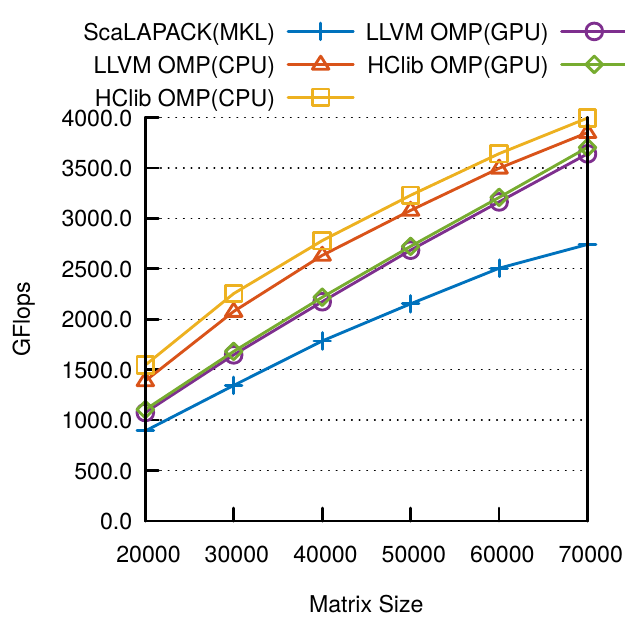}
    \label{fig:perf_lu_multi}
  }\hfill
  \subfloat[QR, 4-node, large size]{
    \includegraphics[width=.22\textwidth, trim=10 5 15 5]{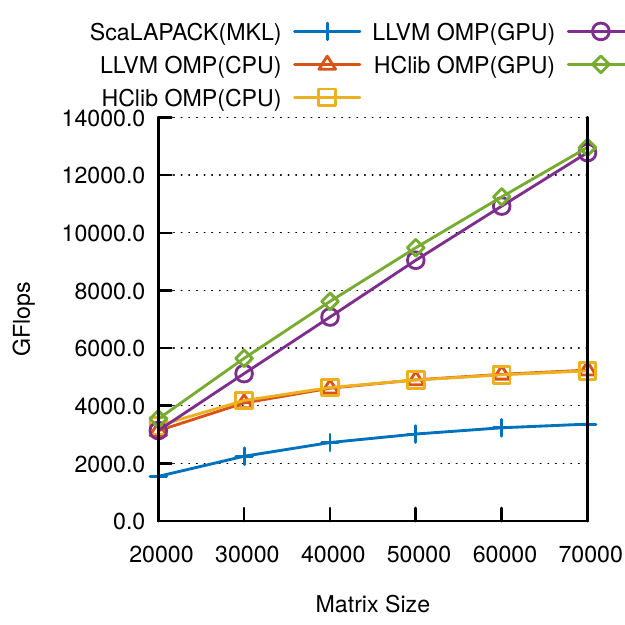}
    \label{fig:perf_qr_multi}
  }
\vspace{-3mm}
\caption{Performance of LU / QR factorization on single / 4-node of
         Cori-GPU (Skylake + V100) with double precision (CPU: CPU-Only, GPU:
         CPU+GPU)}
\label{fig:perf_lu}
%
\end{figure}

Each panel is internally decomposed into tiles. Each thread is
persistently assigned tiles in a round-robin manner, which helps
cache reuse and load balancing. Each thread factors
a column, and an updated trailing matrix in the assigned blocks is
synchronized  at the end of each step (using a custom barrier operation in the library), until a master thread does partial
pivoting across threads and other ranks. Because of these
synchronizations, a user-level threaded runtime without coordination can
lead to deadlock. After the panel factorization, all ranks exchange the
rows to be swapped for partial pivoting; the first rank broadcasts the
top row down the matrix.
The default implementation in SLATE uses a nested parallel region for
the parallel panel factorization. However, this nested parallel region
interrupts the communication and synchronization by oversubscription of
threads on the same cores. Our gang-scheduling makes sure the nested
parallel region runs on reserved workers without  interference
from OpenMP threads in the upper level while other workers can schedule
trailing submatrix tasks for overlapping. As
Figure~\ref{fig:task_graphs} implies, \emph{trailing submatrix
task[i-1]} can run concurrently with \emph{panel task[i]}.
The workers, which are scheduled for gang-scheduling, help to execute
the trailing submatrix tasks by work-stealing when they reach the join
barrier of the nested parallel region.





Figures~\ref{fig:perf_lu_small}, \ref{fig:perf_lu_large}, \ref{fig:perf_lu_multi}
show the performance of LU factorization on single-
and multi-node runs on Cori GPU in double precision. The LU implementation of
SLATE includes the sequential global pivoting phase after the OpenMP
region, so the overall improvement is relatively small compared with
other kernels, which is up to 13.82\% on CPU-only runs. Our
gang-scheduling has diminishing improvement in CPU-only runs with bigger
matrices.
However, with bigger matrices, the GPU version of LU outperforms CPU-only
runs and the reduction in synchronization and communication leads to
noticeable improvement in GPU runs.
We'll explain this performance trend in CPU-only and GPU runs in the
following section.
\begin{figure*}[t]
\footnotesize

  \subfloat[LU] {
    \includegraphics[width=.42\textwidth, trim=10 5 0 15]{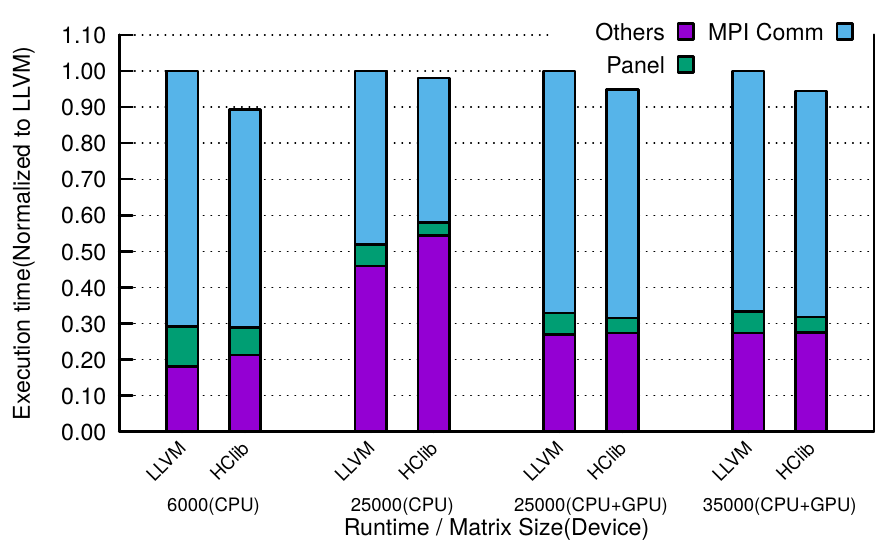}
    \label{fig:stacked_lu}
  }\hspace{1em}
  \subfloat[QR] {
  \includegraphics[width=.42\textwidth, trim=0 5 10 15]{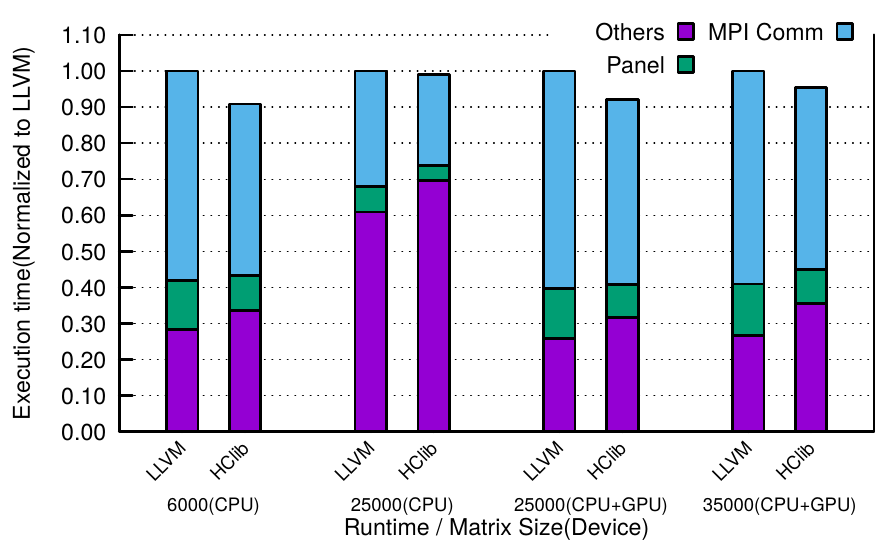}
  \label{fig:stacked_qr}
  }
\vspace{-3mm}
\caption{Detailed Critical Path of LU and QR factorization on a single node with LLVM and HClib OMP}
\label{fig:stacked_lu_qr}
\vspace{-3mm}
\end{figure*}
Similarly, QR factorization does parallel panel factorization.
Unlike LU, QR
doesn't include partial pivoting,
so panel tasks in QR do not involve global
communication for pivoting and QR doesn't have sequential global
pivoting after the parallel region.
Thus, QR factorization shows relatively more significant speed-up with
our runtime over the baseline LLVM OpenMP runtime with oversubscription
compared with LU factorization. SLATE uses a communication-avoiding QR
algorithm for QR factorization. It doesn't include any communication in
the panel factorization, while each panel task transfers the tiles
factored after the panel factorization to other ranks before it proceeds
with lookahead and trailing submatrix tasks. The panel factorization is
also the most critical task to the task graph of QR factorization in
SLATE. Thus, gang-scheduling helps minimize the interference of the
nested parallel regions as it does for LU.

Figures~\ref{fig:perf_qr_small}, \ref{fig:perf_qr_large}, \ref{fig:perf_qr_multi}
show the performance of QR factorization on single- and
multi-node runs. Our work improves the QR factorization up to 14.7\% at
CPU-only runs and 15.2\% at GPU runs on a single node over CPU-only and
GPU runs with LLVM OpenMP runtime. Gang-scheduling shows considerable
improvement in 4-node runs up to 12.8\%.
QR factorization also has diminishing returns of improvement with bigger
matrices, as explained in the following section.

\subsection {Detailed Analysis of Improvement in LU and QR}
\label{sec:lu_qr_anal}
Figure~\ref{fig:stacked_lu_qr} represents 
how much MPI
routines, panel task and other routines consist of the overall execution
time in terms of critical path.
The tasks transfer tiles between ranks in the beginning and end of
panel, lookahead, and submatrix tasks. So, MPI communication and panel
factorization determines the length of the critical path of LU and QR
task graphs. Child tasks from lookahead and trailing submatrix tasks
run in parallel with these routines to overlap the critical routines,
which consists of most portion of \emph{Others}. Each bar is normalized
to the total execution time of LLVM with the corresponding input matrix.




The benefits of gang-scheduling in our integrated runtime for
single- and multi-node runs diminish for both LU and QR factorization.
Gang-scheduling helps remove the delayed synchronization by
oversubscription with deadlock avoidance, which leads to reduction in \emph{Panel}. 
The reduction makes the tile transfer happen earlier, at the end of
the panel task, which shortens the waiting time in
other MPI ranks that need the tiles to proceed. This is shown on the reduction of \emph{MPI Comm} in Figure~\ref{fig:stacked_lu_qr}. 
This improvement is diluted with the combined effect of oversubscription. The degree of degradation incurred by oversubscription depends on the inter-barrier time of an application~\cite{Oversub_MP_IPDPS10}. The bigger input matrix has longer inter-barrier time, which leads to less significant degradation from context switching by oversubscription. Rather, oversubscription hides waiting time from OS and hardware events monitored at the kernel-level, which makes our runtime shows increase in \emph{Others} consisting of single-threaded BLAS kernels. It is because the latency hiding of oversubscription is removed. The decreasing degradation of oversubscription on bigger matrices leads to diminishing returns of gang-scheduling over oversubscription. 

However, the benefit of gang-scheduling becomes more significant on the GPU offloaded version
because a significant portion of computation in \emph{others} is offloaded to
 GPUs where oversubscription helps on the large matrices. 
A larger portion of the single-threaded BLAS kernels is offloaded in LU than in QR. 
So, QR has diminishing returns on the GPU version as the size of the input matrix becomes bigger. 
If more computation in QR is
offloaded, our gang-scheduling can bring more improvement in QR.

\subsection{Cholesky Factorization: Maximized Overlap of Communication and Computation }
\begin{figure}[b]
\vspace{-5mm}
    \centering
     \includegraphics[width=0.37\textwidth, trim=0 0 0 0]{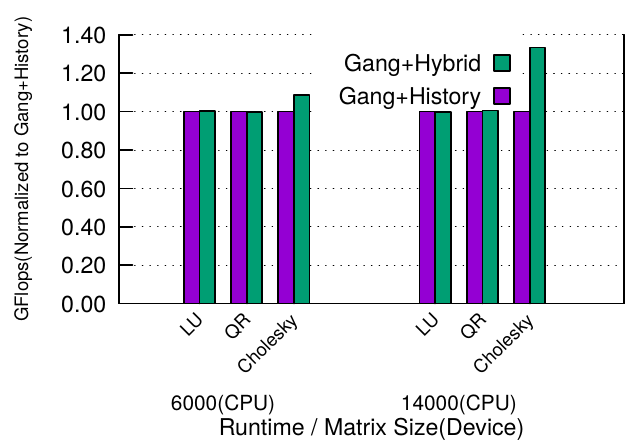}
    \vspace{-3mm}
    \caption{Performance Difference of LU, QR, and Cholesky with history and hybrid victim Selection on HClib OMP}
    \label{fig:hclib_compare}
\end{figure}
\begin{figure}[]
\vspace{-3mm}
     \includegraphics[width=0.45\textwidth, trim=10 10 10 10]{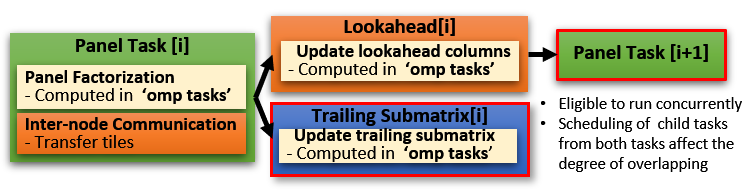}
     \vspace{-2mm}
    \caption{Panel, Lookahead, and Submatrix Tasks of Cholesky in SLATE}
    \label{fig:chol_detail}
\end{figure}
Cholesky factorization is a decomposition of a Hermitian positive
definite matrix into a lower triangular matrix and its conjugate
transpose. Cholesky is used for standard scientific computations such as
linear least squares and Monte Carlo simulations. It has proven to be
twice as efficient as LU when it is applicable. The panel factorization is
much lighter, so lookahead and trailing submatrix tasks are critical to
improving the performance of Cholesky. As we mentioned above,
\emph{trailing submatrix tasks [i-1]} and \emph{panel task[i]} can run
concurrently. LU and QR factorization have heavy \emph{panel tasks} which are
parallelized in a nested-parallel region, so any workers that finish
lookahead tasks will push dependent panel tasks into ready queues. Most
often, they're pushed to the worker's work-stealing queue, so panel
tasks are likely to be scheduled just after lookahead tasks. Also, the
panel tasks are heavy and take a large portion of execution time, so the
degree of overlapping of the panel tasks and trailing submatrix tasks
have limited impact on the performance.  In Figure~\ref{fig:hclib_compare}
, the victim selection policies don't
affect LU and QR significantly while Cholesky is highly influenced by the victim policies which affect the overlapping of the two tasks. 
As described in Figure~\ref{fig:chol_detail}, its panel factorization is done in a bunch of
independent tasks and takes less time than trailing submatrix tasks, so
when the panel task becomes available after its preceding lookahead task
is done, child tasks from the preceding trailing submatrix task are
already being scheduled. The timing for the child tasks from the panel
tasks is determined by how each worker chooses a victim for
work-stealing. If they use the typical history-based victim selection,
every worker will keep stealing from the worker in which the trailing
submatrix is running and create its child tasks. This work-stealing from
the same victim leads to a delay in the scheduling of the panel task and
less overlapping of inter-node communication on the panel task with the
child tasks from the trailing submatrix task.

\begin{figure}[t]
\vspace{-3mm}
\footnotesize

  \captionsetup[subfloat]{farskip=2pt,captionskip=1pt}

  \subfloat[CPU-Only, 1-node] {
    \includegraphics[width=.22\textwidth, trim=10 5 15 10]{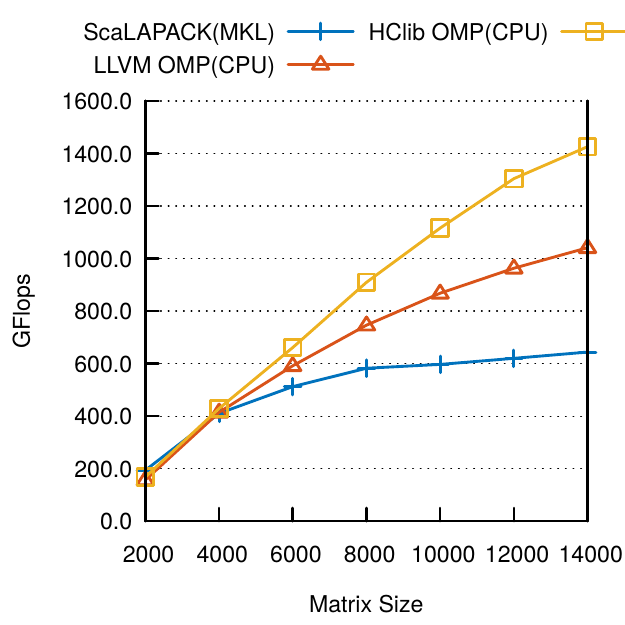}
    \label{fig:perf_chol_small}
  }\hfill
  \subfloat[CPU-Only, 1-node] {
  \includegraphics[width=.22\textwidth, trim=10 5 15 10]{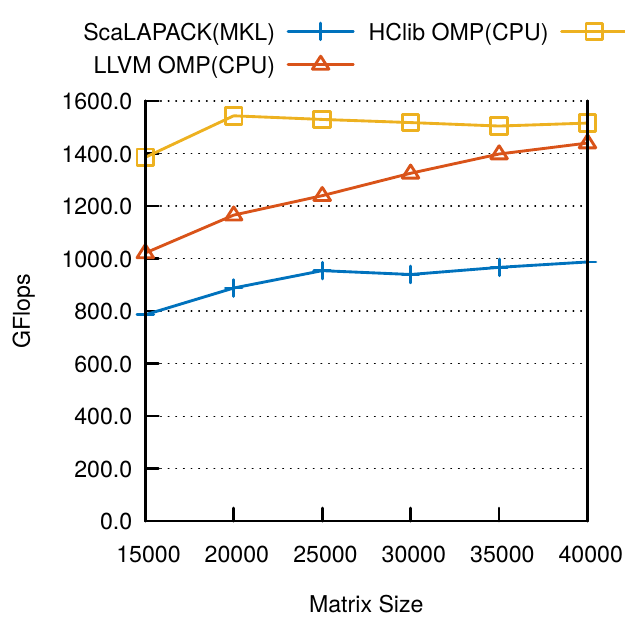}
  \label{fig:perf_chol_large}
  }\vfill
  \subfloat[CPU-Only, 4-node]{
  \includegraphics[width=.22\textwidth, trim=10 5 15 5]{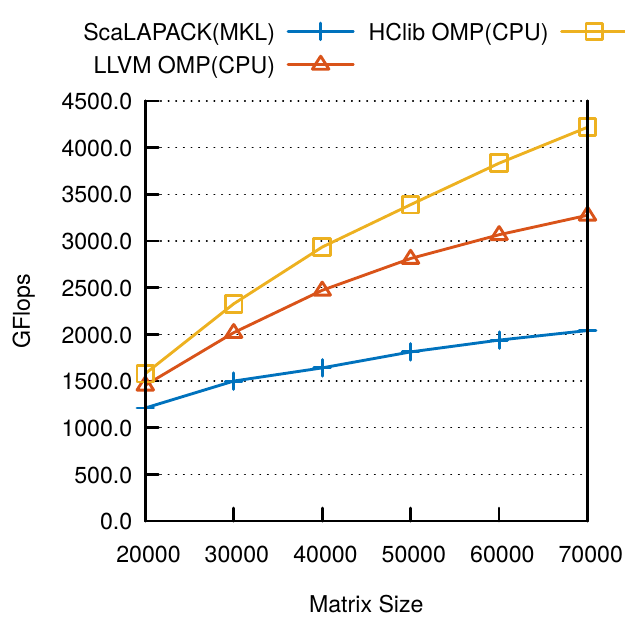}
  \label{fig:perf_chol_multi}
  }\hfill
  \subfloat[Stacked Bar, CPU-Only, 1-node] {
  \includegraphics[width=.22\textwidth, trim=10 5 0 13 5]{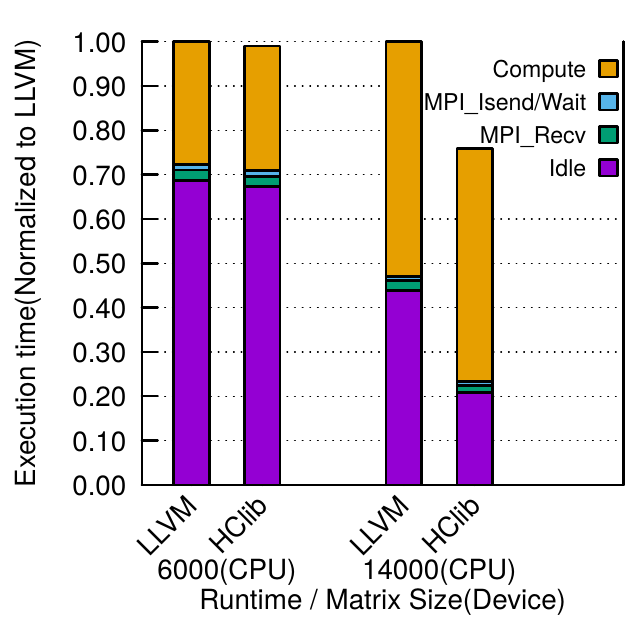} \label{fig:stacked_chol}
  }
\vspace{-2mm}
\caption{Performance of Cholesky factorization on single/4-node of Cori-GPU (Skylake + V100) with double precision (CPU: CPU-Only)}
\label{fig:perf_chol}
\end{figure}
Figures~\ref{fig:perf_chol_small}, \ref{fig:perf_chol_large}, \ref{fig:perf_chol_multi}
show the performance of Cholesky factorization. As we
expected, the improved overlapping of computation in trailing submatrix
tasks and communication in panel tasks enhances the performance of
Cholesky factorization significantly. The improvement is more
significant with bigger matrices because it takes more time to transfer
tiles to other ranks and update the trailing submatrix, which gives more
opportunity for overlapping.
On a single node, the improvement is up to 36.94\% with
double-precision. On 4-node runs, the kernel is improved up to 28.83\%.

We analyze Cholesky in detail to clarify where the improvement comes
from. We profile each OpenMP worker in different MPI ranks and compute
the average of each event such as \emph{Idle}, \emph{MPI\_Recv},
\emph{MPI\_Isend/Wait}, and \emph{Compute} which includes all
computations from panel, lookahead, and trailing submatrix tasks. The largest
portion of \emph{Idle} consists of waiting time until the updated tiles
are received through \emph{MPI\_Recv} from other MPI ranks.
Figure~\ref{fig:stacked_chol} shows the detailed analysis of Cholesky
factorization on a single node with two matrix sizes on LLVM and HClib
OMP. In the small matrix, the amount of computation is relatively small,
which doesn't affect the degree of overlapping significantly regardless
of when MPI routines are called. However, on the large matrix, the
computation from the trailing submatrix takes longer time, which can overlap
MPI routines. So, our victim selection successfully hides the latency of
MPI routines, which leads to significant reduction in the overall idle
time.

\section{Related Work}

\subsection{Task Graphs in Task-Based Parallel Programming Models}

Task graphs have been adopted in most industry and academic works. As
mentioned in earlier sections, languages supporting task graphs provide
constructs for explicit task dependency through objects such as promise
and futures in C++11~\cite{C++11}, Habanero~\cite{barik2009habanero}, Go~\cite{go}. 
A recent work, Legate-Numpy~\cite{Legate}, shows that implicit
parallelism can be extracted from the data flow of library calls.
These task-based parallel programming models supporting task graphs
haven't paid much attention to data-parallel tasks or overlapping of
tasks on the graphs. Hence, we have focused our attention on these
tasks, which are highly crucial for performance.

\subsection{Runtime Systems Based on User-level Threads}
User-level threads have been adopted to benefit from their lightweight
context switching cost. One of the most common uses of ULTs is 
to remove the oversubscription by multiple parallel regions. 
Lithe~\cite{lithe} resolved the
composability of different OpenMP instances by providing a dedicated partition of cores to each instance through user-level contexts. 
However, this partitioning can lead to less resource
utilization because of imbalanced loads across instances. Several runtime systems~\cite{BOLT_PACT19,
ccgrid2018:seon,Aranche,Shenango} share the underlying kernel-level threads through work-stealing or their
own scheduling algorithm with ULTs. They tried to make use of the lightweight
context switching cost of ULTs in different contexts but
couldn't resolve the deadlock issue completely. Shenango~\cite{Shenango}
tried to provide a bypass for blocking kernel calls, but other blocking
operations used in library calls or written by users can lead to a
deadlock.
Our work benefits from the advantages of ULTs without
deadlock or inefficient resource utilization due to coarse-grained
partitioning.


\subsection{Communication and Computation Overlap}
Asynchronous parallel programming
models~\cite{x10:2005,bauer2012legion,kaiser2014hpx,charm++} have been
suggested for overlapping by making all of the function calls
asynchronous, which directs the runtime system to interleave communication
and computation inherently. However, asynchronous parallel programming
models require significant effort on the part of users to write their
applications without deadlock, and tracking control flow of functions
calls is not intuitive. J. Richard et al.~\cite{Euro_par_19_Plasma}
studied the overlapping of OpenMP tasks with asynchronous MPI routines
in which the application uses the \emph{priority} clause and task loops.
As previously mentioned, the examples we used cannot benefit from the
\emph{priority} clause because it works only for \emph{ready} tasks. Our
victim selection helps the overlapping of child tasks from
multiple ready tasks even when \emph{priority} doesn't
help or is not supported. 


\section{Conclusion}
In this work, we proposed gang-scheduling and  hybrid victim selection in our runtime system to improve the performance of task graphs involving inter/intra-node
communication and computation. Our approach schedules nested
parallel regions involving blocking synchronizations and global
communications with minimal interference as well as with desirable data locality. It is implemented efficiently using a monotonic identifier
and an eligibility function to enforce an ordering of gangs so as to ensure the absence of
deadlock. Also, it interoperates with work-stealing to minimize unused
resources within and across gangs.
Our suggested victim selection resolved the problem of the common
heuristic based on a history of previously successful steals by applying
random-stealing and history-based alternatives within a fixed window
size to overlap communication and computation.

We evaluated our work on three commonly used linear algebra kernels,
LU, QR, and Cholesky factorizations, from the state of the art SLATE
library. Our approach showed an improvement for LU of
13.82\% on a single node in double precision and of 11.36\% on multiple nodes.
The improvements for QR went up to 15.21\% on a single node and 12.78\% on
four nodes with double precision. Cholesky factorization was improved by
our hybrid victim selection, with an improvement of up to 36.94\%
on a single node and 28.83\% on multiple nodes with double precision.  Further, unlike current runtimes, our approach guarantees the absence of deadlock in these kernels for all inputs.
Finally, our approach is applicable to any application written using task graphs that also needs to perform additional synchronization and communication operations  as in the SLATE library.


\nolinenumbers
\bibliographystyle{ACM-Reference-format}
\bibliography{paper}
\end{document}